# Nero's 'solar' kingship and the architecture of *Domus Aurea*

**Robert Hannah**

**Faculty of Arts & Social Sciences, University of Waikato, New Zealand**

**Giulio Magli**

**Faculty of Civil Architecture, Politecnico di Milano, Italy**

**Antonella Palmieri**

**Domus Aurea Restoration Project, Rome, Italy**

**Abstract**

*The Domus Aurea, Nero's last "palace" constructed in the very heart of ancient Rome, is a true masterpiece of Roman architecture. We explore here symbolic aspects of the emperor's project, analysing the archaeoastronomy of the best preserved part of the Domus, the Esquiline Wing. In particular, we study the so-called Octagonal Room, the huge vaulted room which is in many respects a predecessor of the Pantheon. The project of the room turns out to be connected with astronomy, as will be that of the Hadrian's Pantheon 60 years later. Indeed, the divinization and "solarisation" of the emperor – placed at the equinoxes as a point of balance in the heavens –are shown to be explicitly referred to in the rigorous orientation of the plan and in the peculiar geometry of the design of the dome.*

# 1. Introduction

Nero is perhaps the most controversial among the Roman Emperors. During his reign (54-68 AD) the empire increased both economically and culturally, but many of his decisions appear to have been inspired by extravagance, if not obsessive tyranny (Champlin 2003a). One of the key aspects of Nero's ideology of power, which appears to have been progressively enhanced by the emperor in the course of his principate, is his close connection with the sun. It is, therefore, natural to investigate whether reflections of Nero's "solar" kingship can be seen reflected in the architecture of the period.

As far as we know, this subject has been previously investigated only by Voisin (1987) who proposed an important hierophany which occurs in the famous Octagonal Room of the Esquiline wing of the Domus. From the architectural point of view, the Octagonal Room is in many respects an antecedent of the Pantheon, being a huge dome with an open circular summit. Recently, two of the authors have provided a  complete re-assessment of the symbolic meaning of the Pantheon in relation to the movement of the sun beam in the course of the year (Hannah and Magli 2011). Two special moments (or better, short periods) are clearly singled out in the Pantheon's project by the archaeoastronomical analysis: (1) the equinoxes, when the sun at noon crosses the apparent spring of the dome and becomes visible from outside through the grille over the entrance, and (2) 21 April, the date of the foundation of Rome, when the sun fully illuminates the entrance. The ideology underlying these hierophanies is linked to the power of Rome – and, by translation, of the emperor – and with the apotheosised emperor himself. Topographical and astronomical links also have been shown to exist between the Pantheon and the mausolea of Augustus and Hadrian, reinforcing the same ideas.

We present here a complete analysis of the same issues for the case of the Esquiline wing of the Domus Aurea, making use also of the results of recent excavations and surveys.

## 2. The Esquiline wing of the Domus Aurea and the octagonal room

The Domus Aurea has had a major influence in western art since its rediscovery in the Renaissance (Dacos 1969). Tentatively, one could apply to it the term *Villa*, i.e. a huge residence with vast porticoes, private apartments, banqueting rooms, and official quarters used for the administration of private and state affairs (Boëthius 1960; Warden 1981; Ball 2003). However, such a Villa was placed in the very heart of Rome, for which reason Nero was censured (Suetonius, *Nero* 31; Tacitus, *Annals* 15.42; Elsner 1994; Davies 2000). The complex covered converging slopes of the Palatine, the Esquiline and the Caelian hills, with an artificial lake placed in the middle. A huge bronze colossus guarded the entrance on the Velia hill (Bergmann 1994,1998; Albertson 2001) **(Fig. 1)**.

The life of the Domus Aurea was very short. The lake was drained by the Flavians for the construction of their Amphitheater, and the statue was moved down near the amphitheater by Hadrian when he ordered the construction of the temple of Venus and Rome (*Historia Augusta*, *Hadrian* 19.12-13; Albertson 2001, Marlowe 2006). Of the Palatine complex very scant traces remain; the Esquiline wing was sectioned by parallel walls, earth-filled and used in the foundations of Trajan's Baths, which today stand over it (Fabbrini 1995, Segala and Sciortino 1999, Ball 2003). The Esquiline wing thus lies today completely underground. It is thispart which is of interest to us here, so we shall describe it in some detail **(Fig. 2)**.

The building originally consisted of two stories. The upper floor – today lost – was probably conceived as a *belvedere* at the foot of the gardens covering the hill. The interpretation of the lower floor has been the subject of much discussion and debate; we adopt here the point of view, backed up by recent excavation work, that it was conceived in two different sections, perhaps chronologically sequential, having different scopes and functions. The section to the west can be truly and effectively compared with a classic suburban *Villa*, characterized by the vast, internal peristyle.This was perhaps a residential sector, devoted to private activities of the emperor. The east section, on the other hand, starts with the service rooms of an open pentagonal court; according to the latest excavations it appears very likely that an identical pentagonal court was placed symmetrically to the east end. The centre of symmetry between the two courts is occupied by the octagonal room, which was therefore the true core of the 'official' part of the Esquiline wing **(Fig. 3)**.

The octagonal room is a masterpiece of Roman architecture: a vaulted dome built on an octagonal basis, one of the first known concrete vaults in the history of architecture (MacDonald 1965; Ball 2003). In a certain sense this room is the precursor of the Pantheon, both from the technical and the symbolic point of view. We shall discuss symbolism later on; from the technical point of view, it is broadly similar but, in many respects, more difficult to construct. The dimensions are as follows: external diameter of the octagon 14.65 m, height of the oculus 9.70 m, diameter of the oculus 5.92 m. The room is, therefore, much smaller than the Pantheon (whose diameter, equal to the maximum height, is 43.3 m); however, the diameter of the oculus with respect to the base of the dome is proportionally much greater in the octagonal room. Furthermore, the architects here managed to cleverly solve a difficult problem: that of smoothing out the octagonal 'arcs' rising from the spring in the homogeneous spherical vault to the top. Since2010 the whole room has been occupied by metal scaffolding for the purpose of restoration, so it was possible to us to examine the vault in detail. The joint between the two geometrical shapes is clearly visible at about 2 meters from the octagonal base of the dome plane (for a complete technical discussion see Giavarini, SamuelliFerretti and Vodret 2000, Conti, Martines and Sinopoli 2009). In the next section, we shall propose a likely explanation of the ideas at the coreof this project.

Another difficulty which was brilliantly solved by the builders is the fact that the vault is deprived of side walls. The spring of the vault is indeed an octagonal perimeter of square bricks (*bipedales*) resting on eight supporting pillars, which serve as jambs for the entrance-ways. As we shall see later in more detail, this allowed the designers to diffuse sunlight coming from 'clerestory' windows located at the level of the extrados of the dome and giving light to the annexes of the room.[1] The oculus was meant to be open to the sky, exactly like that of the Pantheon 70 years later, and a drainage manhole (today occluded) assured the flow of rain water at the center of the room.

A peculiar characteristic of the Esquiline wing is that it was aligned along a rigorous east-west direction. This alignment has been repeatedly noticed by several authors, but – as far as we know – never measured quantitatively; we have thus taken various readings with a precision magnetic compass (½ degree of accuracy once corrected for magnetic declination) along many directions and, in particular, in all the corners of the Octagonal Room; the deviation from cardinality remainseverywhere within one degree. Although the topography of the hill leads naturally to a building which overlooks the valley in a *generic* north-to-south direction, it is difficult to justify in this way such a rigorous cardinal orientation; actually, it would have been more natural to orient the front of the building south of west, in the direction of the lake and of the Palatine hill. Cardinal orientation was by necessity obtained with astronomical observations, either of the sun but more probably of the stars (due to its accuracy). This is quite unusual in the Roman world. Although indeed astronomical orientation (of temples and urban layouts) was perhaps more common than is usually believed (Magli 2008), it does not appear either in the topography of imperial Rome or in the orientation of most imperial temples in Rome. It has, therefore, to be considered as a fundamental ingredient of the design of the Esquiline wing from the very beginning of the project.

In the existing reconstructions, the front of the Esquiline wing is represented as a colonnade, whose extension, however, has never been assessed. Up to recent times it has also never been established whether access to the octagonal room was gained directly from the front or not. In fact, recent trial excavations of the earth-filled south side of the room provided evidence that the main entrance was accessed precisely from the south, through a straight corridor. Traces of the original entrance-way can be seen at the end of this corridor. Together with the perfect symmetry of the east sector with respect to the center of the room, this shows once again that this place really was the focal point of the whole complex. As we shall now see, it was probably conceived with the ideological aim of expressing the 'solarisation' of Nero in tangible, explicit form.

---

[1]   In spite of fanciful reconstructions, the dome was probably not decorated with stuccoes or frescoes (no trace of plaster can be found on it), but the presence of holes hints at wooden, perhaps movable, structures on the ceiling.

## 3. The role of the Sun in the project of the Octagonal Room

In the octagonal room a spectacular hierophany occurs. At local noon on the days of the equinoxes, the annular beam of light entering the oculus hits the jamb of the northern entrance-way. The height of the upper opening and its width are therefore calibrated in such a way that the lower rim of the sun beam strikes the juncture between the floor and the doorway's threshold, while the beam encircles the jambs. So the sun measures out the dimensions of this opening, and by extension of the other openings, in the walls of the room.

The existence of this phenomenon is not well recognised: it was apparently discovered by Robert Marchand and Jerome Veyrin, who, however, left it unpublished. They,however, documented it with a striking picture later published by Voisin (1987) and reproduced here (**Fig. 4**). Unfortunately, it was (and is) impossible for us to document it afresh, because the oculus is covered by an immovable, opaque glass protection. It has to be hoped that once the restoration of the monument is completed, the protection of the oculus might be removed and substituted by a clear glass screen, so that the sun can once again be on time at its appointment fixed two millennia ago, twice a year at the equinoxes. For the time being, we have performed an accurate computer simulation (using laser-scanner measurements and the software Starry Night Pro 6.0) of the movement of the sunlight inside the room at the time of construction and today (the latter positions are very slightly displaced due to the variation of the obliquity of the ecliptic), in order to control the equinox phenomenon and to check for other possibly relevant configurations.

The results of such a simulation, together with the new data on the entrance-ways, leave little doubt that the whole project was governed from the very beginning by the wish to obtain the spectacular equinox hierophany. Indeed, the calculated altitude of the sun at the equinoxes in Rome in 64 BC is $\alpha=47° 30´$, which coincides almost perfectly with the angle formed by a line connecting the spring of the dome at the doorway with the rim of the oculus (today the sun's altitude is only slightly lower, about 46° 56´, so that the phenomenon is still visible) (**Fig 5**). The fact that, due to the recent explorations, we can today be sure that the main entrance to the room was sited in a south-to-north direction from the front of the building strongly supports the non-accidental nature of the phenomenon; indeed, as we shall see, the 'solarised' Nero was connected with the sun at the equinoxes in contemporary literature, and it appears likely that what took place in the room was a sort of 'imperial hierophany'. The northern room, inside which the beam of light enters like the light of a modern reflector, was occupied by a fountain, or nymphaeum. People entering the room at or near noon would have perceived the corresponding ideology 'actualized' by the positioning of the sun beam inside the room. Due to the fact that the azimuths and altitudes of the sun at days near the equinoxes vary with great rapidity, the phenomenon was perceptible only for a few days.

Inspired by the corresponding analysis we made for the Pantheon, we further performed a full simulation of the movement of the sun beam inside the room in the course of the year. The complete results are as follows (dates in Gregorian calendar, but in Nero's time the delay of the Julian calendar was of course imperceptible).

1) The sun wanders on the domed surface approximately between November 6 and February 10, effectively through the season of winter, which for instance Varro (*de Re Rustica* 1. 28. 1-2) sets between 10 November and 7 February.

2) In mid-February the sun leaves the base of the dome at noon and starts entering the northern room at local noon. In the first week of March the sun directly illuminates a person standing in the doorway – the timing corresponds to the period, 5 March, when Nero became pontifex. The symmetric day is around the accession of Nero, on 13 October. Both dates, 5 March and 13 October, were formally commemorated by the Arval Brethren in Nero's reign (Scheid 1998), so they are part of the 'cultural baggage' which occupants of the room could have brought with them, enabling them to recognise the solar associations of the room at these times of the year.

3) As mentioned above, the northern room is fully lit at the spring equinox.

4) Later in the year, the sun starts illuminating the north-east/north-west rooms (which are fully illuminated when the sun has azimuths 135° and 225°). They receive the full beam around 14 April. In the days bridging the summer solstice up to 3 September, when the side rooms will be fully lit again, the sun still partially illuminates them, while the east and west rooms are never fully illuminated. In the same period, at noon, the sun beam hits the floor, reaching (of course) the point nearest to the center at the summer solstice.

5) Interestingly, only for some ten days before and after the summer solstice was the direct light of the sun (culminating at ~70° 40´) able to reach the floor of the northern room from the 'clerestory' windows, all of which are angled at 68°. On all other days they furnished only a sort of suffused illumination to the sides of the main ambient.

These characteristics, and particularly the equinoctialhierophany, suggest astronomy was a key part of the project. To check this we preliminarily observe that the room was constructed in accordance with a refined aesthetic, which is quite different from – and in a sense more complicated than – the one followed in the Pantheon some 60 years later. In the Pantheon, indeed, the height of the base cylinder is equal to the radius of the dome, and, as a consequence, an ideal, huge sphere perfectly fills the interior. A person standing at the center of the floor is also at one of the "poles" of this sphere, the other one being located in the ideal completion of the sphere in the center of the upper opening. In the octagonal room, on the other hand, the height of the octagon "drum" (the analogue of the Pantheon's cylinder) is ideally equal (actually 4.91 m as opposed to 4.79 m) to the height of the rim of the oculus with respect to the upper "drum" base. The construction of the dome on the base of the octagonal spring clearly was a very complicated engineering problem, solved – as mentioned before – with a sort of "smoothing" procedure: concrete was led down using octagonal wooden shapes, turned in a circular *centina* at about 2 meters. Due to this fact, and since the diameter of the oculus is quite wide, the radius of the upper, truly spherical part of the dome is a lot greater than half the height (7.64 m if calculated along the apexes' section). Geometrically, this means that the center of curvature of the dome is located at a virtual point, lower than the spring height but well above the floor; consequently, the room gives to the visitor an impression of 'floating space', which is thus quite different frombut not less fascinating thanconveyed by the Pantheon.To reconstruct the way in which the architects of the octagonal room probably planned such a refined project we make reference to an "idealization" of it outlinedin**Fig. 6.** First of all, the angle α was calculated by accurate solar observations of the shadow of a stick of known height; actually, for the whole construction which follows it is sufficient to work with the co-tangent of α(the ratio between the two sides of a squared triangle having α as one of the angles) which we shall denote by k. Then, the height of the drum – the segment AB in Fig. 6 – was chosen. Consequently, the width of the oculus A′B′ was fixed (as it is k times AB). At this point, what remained free was the height of the oculus, and an aesthetic canon was selected: to render it (theoretically) equal to the height of the drum. As a consequence of such choices, simple geometry shows that the height h of the centre of curvature of the dome – that is,the height of the centre of the arc of circumference connecting B and B′ - is fixed by the two parameters AB and k,ash=AB (3/2 –k^2). Thus,the centre of curvature of the dome would be located precisely at half the height of the drum for $\alpha$ =45° (k=1) and it is actually located a bit higher than this value (2.70 meters compared to 2.30) as $\alpha$ is slightly greater than 45°.

## 4. Discussion

To understand the role of the equinoctialhierophany in Nero's ideology of power, we first of all observe that – as is well known –the connection between the divinized emperor and the celestial realm was already established by Augustus at Julius Caesar's death with the help of the appearance of a comet. Caesar's catasterism becomes explicit in Ovid's *Metamorphoses*, written ca. AD 8, where Caesar's soul 'shines as a star' (Ovid, *Metamorphoses* 15.850). Other relevant passages suggest that a particular point of cosmic balance was understood to be assigned to the emperor in the heavens. For instance, Vergil (*Georgics* 1. 24-35) assigns a Caesar to a place in the heavens between Virgo and Scorpius; slightly later Manilius(*Astronomica* 4. 546-551, 773-777), writing early in Tiberius's reign, places Caesar – this time probably Tiberius himself – in Libra (Lewis 2008). Libra, the Balance, at that time at the autumn equinoctial point, is a point of balance between the ecliptic and the equator. Finally, according to various sources, Nero raised himself to a personal identification with the Sun, and therefore to a sort of divinization of the *living* emperor (L'Orange 1947; Hiesinger 1975; Varner 2000). The extent, genesis, and development of such an identification are the subject of current debate (Bergmann 1994, 1998; Smith 2000; Albertson 2001; Pollini 2010). Reasonably, however, the first date when the idea of a new, Apollonian golden age under the 'Solar emperor' starts to be promoted is AD 59, as is discernible in Lucan's poetry (*Pharsalia*1.45–59) and in coinage depicting Nero as the new Apollo (Hiesinger 1975; Bergmann 1998; Albertson 2001; Champlin 2003a). It is thus contemporary with the conception of the Domus Aurea, when the ideology seems to reach its apex, with Nero identified with, or at least imaged in the form of, Sol, the benefactor of mankind (Bergmann 1994, 1998; Smith 2000; Pollini 2010). In this context the coronation of Tiridates in the Roman forum in 66 is suggestive. Nero was sitting on the Rostra, fronting the Via Sacra (thus looking east/south east), and therefore facing the rising sun – Champlin (2003a) imagines Nero's face lit by the sun – and Tiridates pronounced the formula 'I have come to you, my god, to worship you as I do Mithra.' (Dio Cassius 63.5.2) As is well known, the solar connotations of Mithra were very strong (Champlin 2003a; Beck 2004, 2006).

It should be noted that, within the context of Nero's life, there is no apparent special significance in theequinoctial dates. In his own time the Arval Brethren commemorated several anniversaries associated with the emperor, some of which we have already noted: 25 February as the date of his adoption; 5 March when he became pontifex; 11 September for his safety (*salus*) and return (*reditus*) – from what is not stated; 13 October when he gained *imperium* and acceded to the throne; 4 December when he gained tribunician power; and 15 December as his *Dies Natalis* (Scheid 1998). The spring equinox, however,would lie near to the time of Nero's supposed conception and, as is well known, Augustus – born near the autumn equinox – definitively chose in many instances to associate his image with that of the zodiac sign Capricorn, the sign of the winter solstice and therefore of the date of his conception (Dwyer 1973, Zanker1988). In addition, in Neroian poetry the 'proper seat' of the emperor is the celestial equator. This is confirmed by Lucan (1.45-59), writing in ca. 60AD, who has the apotheosized emperor joining the heavens and finding his final place on the celestial equator, where he will ensure balance and stability.

To summarize, and without wishing to stretch things so far as to define the Domus Aurea a sort of 'house of the living sun' (Charlesworth 1950), we believe that a role of the palace in exploiting the solar ideology is by no means surprising. This was perhaps also enhanced by the topography of its main elements; indeed, although the exact position of the Colossus over the Velia hill is unknown, the area where it was located before Hadrian's relocation cannot exceed the perimeter of the temple of Venus and Rome. Using satellite images (see again **Fig. 1)** we can therefore deduce that the statue was located at an azimuth which can span from a few degrees south of west to true west when seen from the terrace of the Esquiline wing. Interestingly enough, this means that the head of the Colossus was seen, by an observer looking from the Esquiline wing, suddenly illuminated by the rising sun, in the days near the equinoxes, perhaps corresponding to Suetonius' words about the birth of Nero, illuminated by the first rays "*exoriente sole*."

All in all, then, Nero's octagonal room is – like the Pantheon and actually before it – a convincing case in which architectural elements in a Roman building were symbolically tied to the sun's cycle, acting as cosmological signposts for those Romans sensitive to such things.

**Acknowledgements**

The authors wish to thank the authorities of the Soprintendenza Speciale per i BeniArcheologici di Roma for allowing them access to the monument.

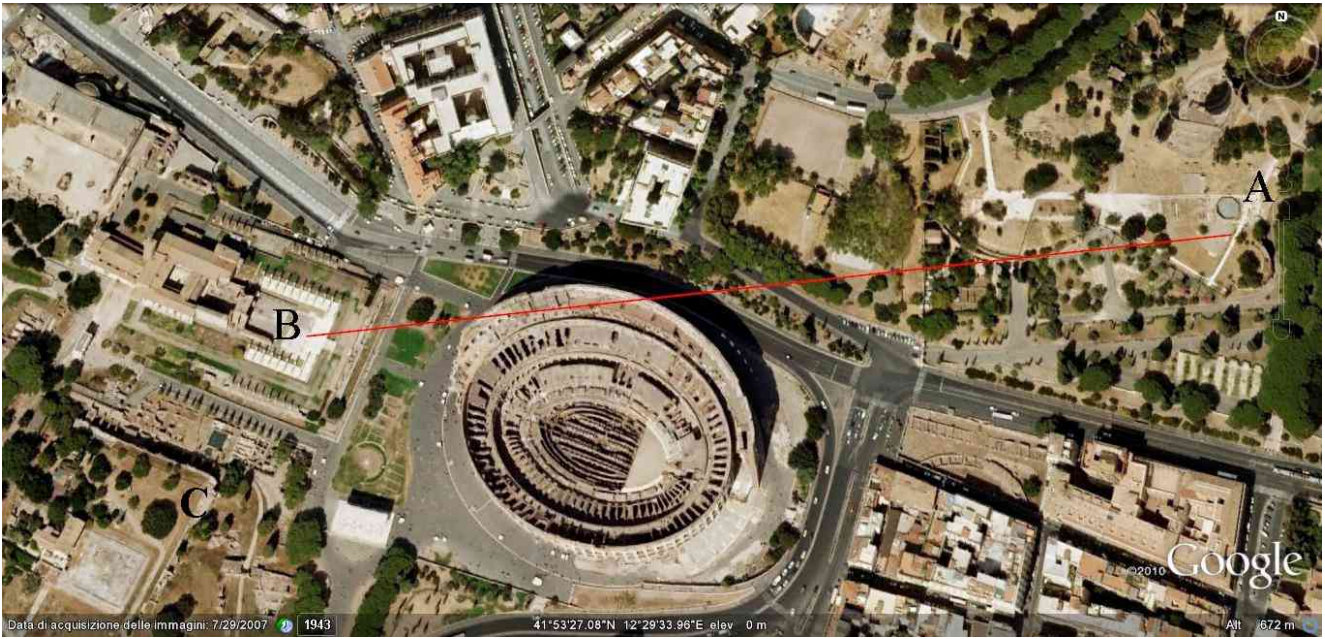

**Fig. 1**

**Satellite image of Rome, showing the area of the Domus Aurea.**

**A) Esquiline Wing (only the grey octagon which is the modern covering of the Octagonal room is visible) B) Velia sector, in the area of the temple of Venus and Rome. The letter denotes the likely location of the Colossus. C) Area of the recent excavations on the Palatine Hill. The Colosseum occupies the area of the lake. The red line showing the relationship between the Esquiline wing and the Colossus area bears an azimuth of approximately 264°. (Image courtesy Google Earth, drawings by the authors)**

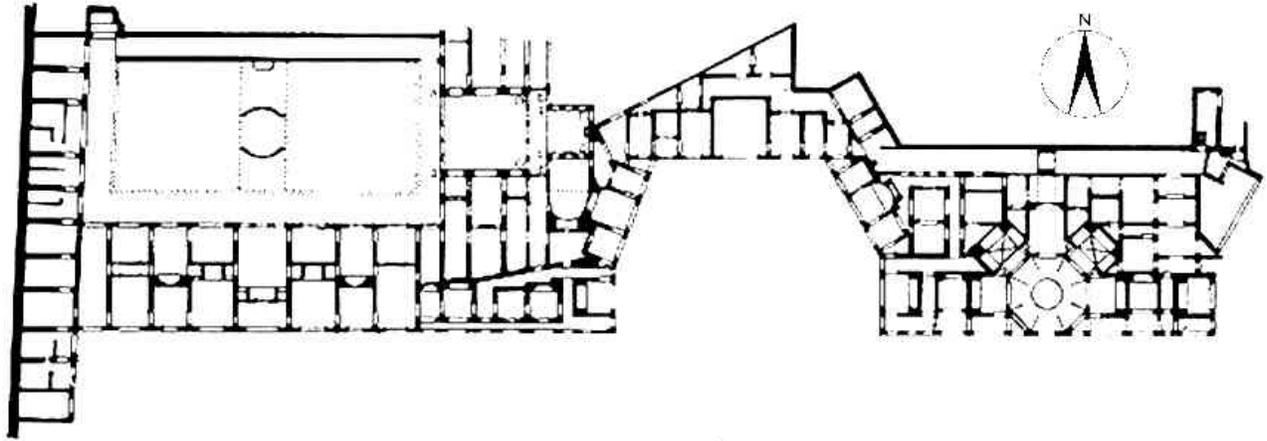

**Fig. 2 Plan of the Esquiline wing of the Domus Aurea.**

**To the left (west) the private part, organized as a Roman Villa. Then the public part, with the first polygonal court (center) and the octagonal room; with all probabilities a symmetric court (un-excavated) was placed to the right so that the octagonal room was the balance point of the whole public part of the complex. In front (south) of the octagonal room an entrance corridor was located (image by the authors, redrawn from various sources)**

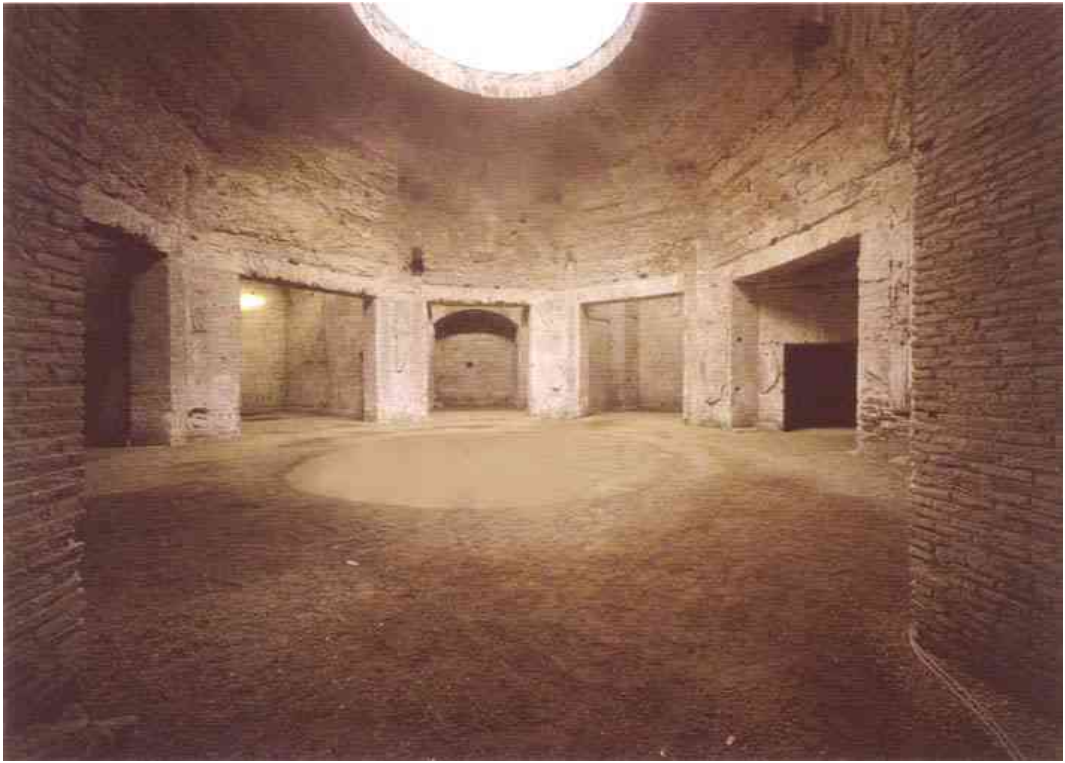

**Fig. 3. The octagonal room (Archivio SSBAR, foto E. Monti, Permission requested,)**

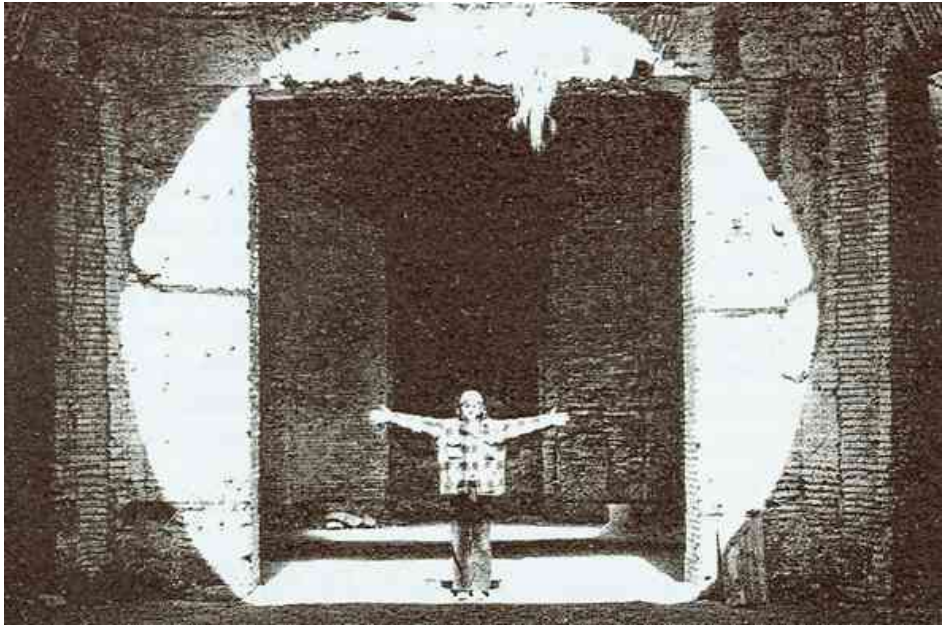

**Fig. 4 The equinox hierophany in the octagonal room, documented for the first time in a picture by Robert Marchand, published by J-L. Voisin.**

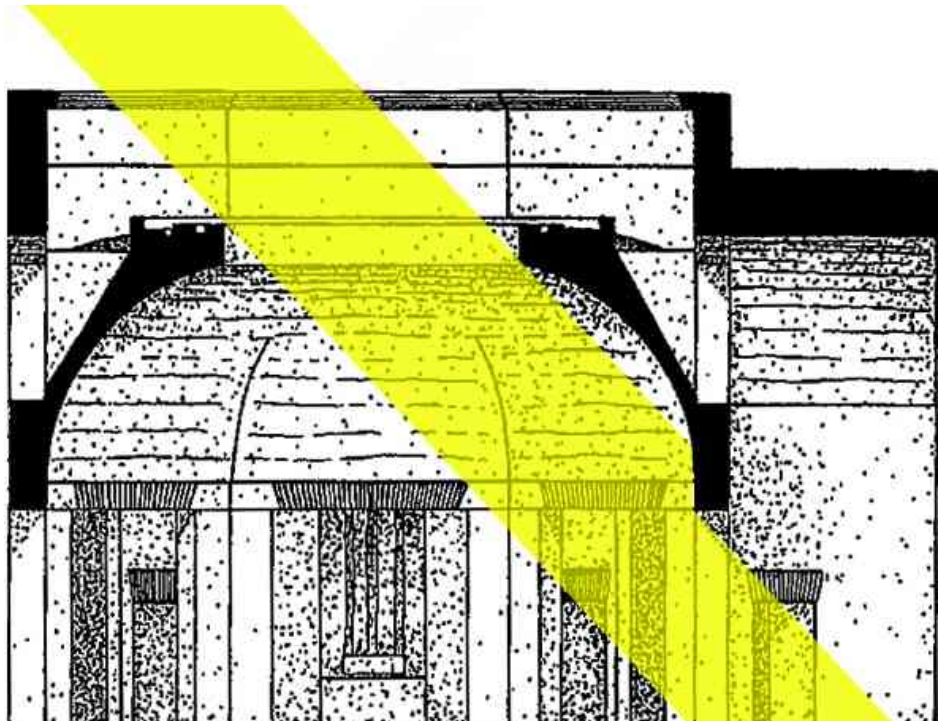

**Fig. 5**

**North-south section through the Octagonal Room, showing the fall of the noon sunlight at the equinoxes  (drawing Robert Hannah, after Ball 2003: 210 fig. 72 with permission).**

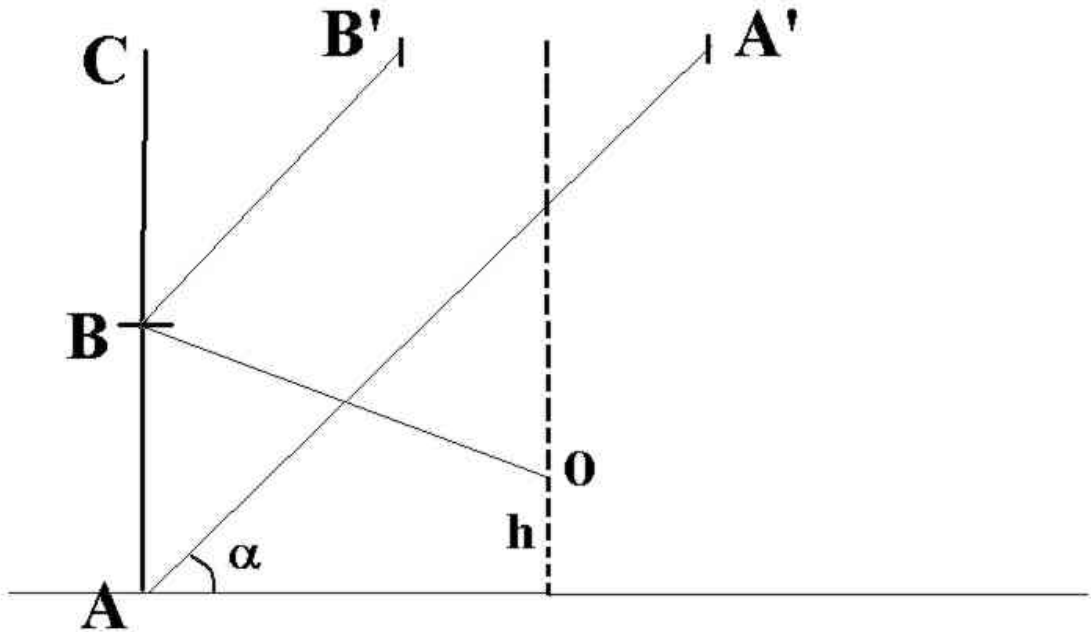

**Fig. 6**

**A schematic, ideal reconstruction of the "solar" geometry at the basis of the Octagonal Room's project (the reader should be warned that the actual measurements of the room are only roughlyrepresented inthis construction; the section is drawn through the apexes): $\alpha$ altitude of the sun at noon at the equinoxes; AB height of the drum, AC=2 AB height of the oculus, A′B′ width of the oculus, O center of the dome (of radius OB) connecting B and B′, h height of the center of the dome.**